 \documentclass{jetpl}
\twocolumn
\title{Topological invariants  for Standard Model: from semi-metal to topological insulator}

\rtitle{Topological invariants  for Standard Model: from semi-metal to topological insulator}

\sodtitle{Topological invariants  for Standard Model: from semi-metal to topological insulator}

\author{
G.E. Volovik
 \/\thanks{
volovik@boojum.hut.fi} 
}

\rauthor{
G.E.Volovik
}

\sodauthor{
Volovik
}

\address{Low Temperature Laboratory, Helsinki University of
Technology, P.O.Box 5100, FIN-02015, HUT, Finland
\\
 Landau Institute for Theoretical Physics RAS, Kosygina 2,
119334 Moscow, Russia}

\dates{December 13, 2009}{*}

\PACS{ }
\abstract{
We consider topological invariants describing semimetal (gapless) and insulating (gapped) states of the quantum vacuum of Standard Model and possible quantum phase transitions between these states. 
}

\begin{document}

\maketitle


\section{Introduction}

Recently topological insulators, semimetals, superconductors, superfluids and other topologically nontrivial gapless and gapped phases of matter have attracted a lot of attention. 
Probably the first discussion of the 3D topological insulators in crystals can be found in Refs. \cite{Volkov1981,VolkovPankratov1985}; the  two-dimensional massless edge states of electrons at the
interface between topologically different bulk states have been discussed in \cite{VolkovPankratov1985}. The fully gapped 3D superfluid with nontrivial topology is represented by the phase B of superfluid $^3$He; 
the corresponding 2D gapless quasiparticles living at interfaces have been discussed in  \cite{SalomaaVolovik1988}. Examples of the  2D  topological fully gapped systems are provided by the 
films of superfluid $^3$He in the phase A and in the planar phase; the topological invariants give rise to  quantization of the Hall and spin-Hall currents in these films in the
absence of external magnetic field  \cite{VolovikYakovenko1989}. The three-dimensional $^3$He-B and the two-dimensional planar phase of triplet superfluid/superconductor belong to the time-reversal invariant topological states of matter. 

Different aspects of physics of topological matter have been discussed, including topological stability of gap nodes; classification of fully gapped vacua;
edge states; Majorana fermions; influence of disorder and interaction; topological quantum phase transitions; intrinsic Hall and spin-Hall effects; quantization of physical parameters; experimental realization; connections with relativistic quantum fields; chiral anomaly; etc. 
\cite{VolovikMineev1982}-\cite{Ryu2009}

 The vacuum of Standard Model (SM)  is also  a  topological substance: both known states of the SM vacuum -- gapless semimetal state and fully gapped insulating state -- possess  the non-trivial topological invariants. Here we present the explicit expression for the relevant topological invariants of SM, and discuss possible topological quantum phase transitions occurring between the vacuum states.

\section{Green's function as an object}

The object for the topological classification must be the Green's function rather than Hamiltonian. Then it is applicable even in cases when one cannot introduce the effective low energy Hamiltonian, for example when  Green's function does not have poles, see \cite{Volovik2007,FaridTsvelik2009}
in condensed matter,
 unparticles in relativistic quantum fields \cite{Georgi2007} and phenomenon of quark confinement  in QCD with suggested anomalous infrared behavior of the quark and gluon Green's functions \cite{Gribov1978,Chernodub2008,Burgio2009}.

Green's function topology has been used in particular  for classification of topologically protected nodes in the quasiparticle energy spectrum of systems of different dimensions
including the vacuum of Standard Model in its gapless state \cite{FrogNielBook,Volovik2003,Horava2005,Volovik2007}; for the classification of the topological ground states in the fully gapped $2+1$ systems, which experience intrinsic quantum Hall and spin-Hall effects  \cite{VolovikYakovenko1989,Yakovenko1989,SenguptaYakovenko2000,ReadGreen2000,Volovik2003,Volovik2007};   in relativistic quantum field theory  of $2+1$ massive Dirac fermions 
  \cite{IshikawaMatsuyama1986,IshikawaMatsuyama1987,Jansen1996}; etc.

The quantum phase transition occurs when some parameter of the system crosses the critical value at which the momentum-space topology of the Green's function changes. In SM the role of such parameter may be played by the high-energy cut-off scales, such as the ultraviolet $E_{\rm UV}$ and compositeness $E_{\rm c}$ energy scales introduced in Ref. \cite{KlinkhamerVolovik2005a}. In the limit $E_{\rm UV}/E_{\rm c} \rightarrow \infty$, all three running coupling constants vanish, $(\alpha_1, \alpha_2,\alpha_3) \rightarrow 0$ \cite{KlinkhamerVolovik2005a}. In this zero-charge limit  fermions become uncoupled from the gauge fields, the gauge invariance becomes irrelevant, and the gauge groups of SM may be considered as the global groups which connect the fermionic species.
That is why in this limit the Green's function is well defined. 

In general case when the gauge invariance is important there are two different approaches to treat the Green' function. (i) One may use the gauge-fixing conditions. Though in this approach 
the Green's function depends on the choice of the gauge,  the topological invariants are robust to continuous deformations and thus should not depend on the choice of the gauge (only large gauge transformations are prohibited since they may change the value of topological invariant).  (ii) One may use the two-point Green's function with points connected by a special path-ordered exponential, which produces the parallel transport of color from one point to the other and thus makes the propagator gauge invariant. In this case the Green's function is path-dependent, and we assume the straight contours to ensure translational invariance, which allows us to consider the Green's function in momentum space.  

Simplest examples of the Green's function are $2\times 2$ matrix Green's function for chiral Weyl fermions 
\begin{equation}
S =\frac{Z(p^2)}{i\omega \pm \mbox{\boldmath$\sigma$}\cdot{\bf p}}~,
\label{Weyl}
\end{equation}
with $+$ sign for right and  $-$ sign for left fermions, and 
 $4\times 4$ matrix Green's function for Dirac fermions:
\begin{equation}
S  =\frac{Z(p^2)}{-i\gamma^\mu p_\mu+ M(p^2)}~.
\label{Dirac}
\end{equation}
Here ${\mbox{\boldmath$\sigma$}}$ are spin Pauli matrices; Dirac matrices will be chosen according to Sec. 5.4 in Ref. \cite{WeinbergQTF}:
  \begin{equation}
  \gamma^0=-i\tau_1~~,~~ {\mbox{\boldmath$\gamma$}}= \tau_2{\mbox{\boldmath$\sigma$}}~~,~~
  \gamma_5=-i\gamma^0\gamma^1\gamma^2\gamma^3=\tau_3
 \,.
\label{eq:DiracMatrices}
\end{equation}
For the topological classification of the gapless vacua, the Green's function is considered at imaginary frequency $p_0=i\omega$, i.e. the Euclidean propagators are used 
and
 \begin{equation}
 p^2={\bf p}^2 -p_0^2={\bf p}^2 +  \omega^2 \,.
\label{eq:p^2}
\end{equation}
This allows us to consider only the relevant singularities in the Green's function and to avoid the singularities on the mass shell, which exist in any vacuum, gapless or fully gapped.

There is the following topological invariant expressed via integer valued integral over the $S^3$ surface $\sigma$ around the point $p^2=0$ in momentum space \cite{Volovik2003}:
 \begin{equation}
N = \frac{e_{\alpha\beta\mu\nu}}{24\pi^2}~
{\bf tr}\int_\sigma   dS^\alpha
~ S\partial_{p_\beta} S^{-1}
S\partial_{p_\mu} S^{-1} S\partial_{p_\nu}  S^{-1}\,.
\label{MasslessTopInvariantSM1}
\end{equation}
Here trace is over all the fermionic indices of the Green's function matrix including spinor indices.
This invariant \eqref{MasslessTopInvariantSM1} equals the difference between numbers of right and left fermionic species, 
 \begin{equation}
N =n_R - n_L\,.
\label{MasslessTopInvariantvalue}
\end{equation}
One has $N=+1$  for a single right fermionic species in \eqref{Weyl};  $N=-1$ correspondingly for a single  left fermion. For massless Dirac fermions  one has $N=+1-1=0$, which demonstrates that there is no topological protection, and interaction may produce mass term in \eqref{Dirac}. Eq.\eqref{MasslessTopInvariantvalue} is applicable to the general case when interaction between the fermions is added, or Lorentz invariance is violated. Nonzero value $N$ of the integral around some point in momentum space tells us that at least $N$ fermionic species  are gapless and have nodes in the spectrum at this point.

\section{Topological invariant protected by symmetry in semi-metal state}
\label{TopSemi-metal}

We assume that SM contains equal number  of right and  left Weyl fermions, $n_R =n_L =8n_g$ , where $n_g$  is the number of generations (we do not consider SM with Majorana fermions, and assume that in the insulating state of SM neutrinos are Dirac fermions).
For such Standard Model the topological charge in \eqref {MasslessTopInvariantSM1} vanishes, $N=0$. Thus the masslessness of the Weyl fermions is not protected by the invariant \eqref{MasslessTopInvariantSM1}, and arbitrary weak interaction may result in massive particles.

However, there is another topological invariant, which takes into account the symmetry of the vacuum. The gapless state of the vacuum with $N=0$ can be  protected by the following
integral   \cite{Volovik2003}:
 \begin{equation}
N' = {e_{\alpha\beta\mu\nu}\over{24\pi^2}}~
{\bf tr}\left[K\int_\sigma   dS^\alpha
~ S\partial_{p_\beta} S^{-1}
S\partial_{p_\mu} S^{-1} S\partial_{p_\nu}  S^{-1}\right]\,.
\label{MasslessTopInvariantSM}
\end{equation}
where $K_{ij}$ is the matrix of  some symmetry transformation. 
In SM there are two relevant symmetries, both are the  $Z_2$ groups, $K^2=1$. One of them is the center subgroup of $SU(2)_L$ gauge group of weak rotations of left fermions, where the element $K$ is the gauge rotation by angle $2\pi$, $K=e^{i\pi {\check \tau}_{3L}}$. The other one is the group of the hypercharge rotation be angle $6\pi$, 
$K=e^{i6\pi Y}$. In the  $G(224)$ Pati-Salam extension of the $G(213)$ group of SM, this symmetry comes as combination of the $Z_2$ center group of the $SU(2)_R$ gauge group for right fermions, $e^{i\pi {\check \tau}_{3R}}$, and  the element $e^{3\pi i(B-L)}$ of the
$Z_4$ center group of the $SU(4)$ color group -- the $P_M$ parity (on the importance of the discrete  groups in particle physics see 
\cite{PepeaWieseb2007,Kadastik2009} and references therein).
Each of these two $Z_2$ symmetry operations   changes sign of left spinor, but does not influence the right particles. Thus these matrices are diagonal, $K_{ij}={\rm diag}(1,1,\ldots, -1,-1,\ldots)$, with eigen values 1 for right fermions and  $-1$ for left fermions.

In the symmetric phase of Standard Model, both matrices commute with the Green's function matrix $S_{ij}$, 
as a result $N'$ is topological invariant: it is robust to deformations of Green's function which preserve the symmetry.
Simple explanation is the following.
The $Z_2$ symmetries $K$ completely forbid the  mixing, $M(p^2)=0$, and one has two independent sectors in the Green's function matrix with two independent topological invariants $N=N'=n_R$ for  right fermions and $N=-N'=-n_L$ for left fermions. Thus the symmetric phase of Standard Model contains  $N'=16n_g$ massless fermions, which remain massless even if the interaction is introduced. 
Since the mixing between leptons and quarks is negligibly small at low energies, the 
invariant \eqref{MasslessTopInvariantSM} splits into two separate invariants for leptonic and baryonic sectors $N'=N'_{\rm leptons}+ N'_{\rm baryons}=4n_g + 12 n_g$.

This is one of numerous examples of  integer valued  topological invariants which are supported by   discrete or continuous symmetry. Other examples in condensed matter systems and in quantum field theory can be found in Refs. \cite{VolovikYakovenko1989,Volovik1992b,SenguptaYakovenko2000,Volovik2003,Volovik2007}.

 \begin{figure}[top]
\centerline{\includegraphics[width=1.0\linewidth]{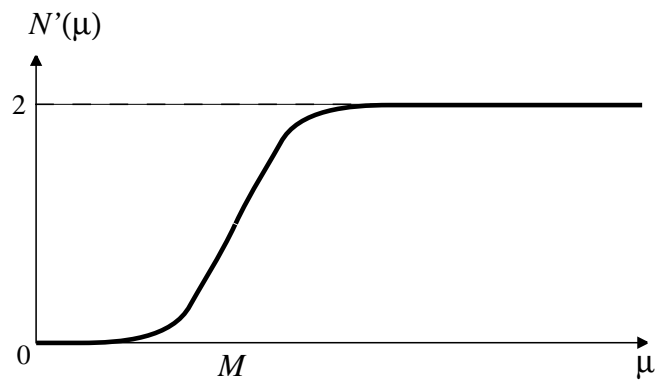}}
  \caption{Fig. 1. Integral \eqref{MasslessTopInvariantSM} as function of running energy $\mu$ in the vacuum  of a system with one right and one left fermions  in the insulating  and  symmetric gapless states. (i) In the symmetric state (dashed line), equation  \eqref{MasslessTopInvariantSM} represents the topological invariant protected by symmetry $K$. Its magnitude $N'$ does not depend on the choice of the 3D surface $\sigma$. (ii) In the insulating state  (solid line)  there is no symmetry, which could protect the topological invariant. That is why  fermions become massive Dirac particles.  The operator $K$ does not commute with the Green's function, as a result, the integral $N'$ in \eqref{MasslessTopInvariantSM} is not  invariant under deformation of the 3D surface $\sigma$ and depends   on its radius $\mu$ of the $S^3$ sphere:  $\omega^2+{\bf p}^2=\mu^2$. The function $N'(\mu)$ experiences a crossover at $\mu\sim M$,   where  $M$ is mass  of Dirac particle. Similar crossovers  between plateaus with different integer value of the integral \eqref{MasslessTopInvariantSM} occur in SM.}
 \label{Crossover} 
\end{figure}

The integral \eqref{MasslessTopInvariantSM} can be applied also to vacua with massive Dirac fermions. In the massive case this integral is no more the topological invariant:  it depends in particular on the running mass parameter $\mu$ -- the radius  of the $S^3$ sphere $\sigma$ about the origin, ${\bf p}^2+\omega^2=\mu^2$. The function $N'(\mu)$ for the Dirac vacuum is illustrated in Fig. \ref{Crossover}.  In the limit of large  $\mu \gg M$, the interaction between right and left fermions vanishes, the symmetry  $K$ is restored, and the integral approaches the  value  $N'=2$. In the opposite limit, $\mu \ll M$, the intergral vanishes, because there is no symmetry  which may protect the singularity at $p^2=0$.

\section{Topological invariant protected by symmetry in insulating vacuum}

In the asymmetric phase of SM, there is no mass protection by topology and all the fermions are massive, i.e. SM vacuum becomes the fully gapped insulator. In principle the transition between the gapless vacuum of SM and its insulating state could be the topological quantum phase transition as in condensed matter 
(see \cite{Volovik1992b,KlinkhamerVolovik2005b,GurarieRadzihovsky2007}). 
There are also the other possible quantum phase transitions -- topological transitions between different insulating states with different  topological charges relevant for the fully gapped fermions.

In the fully gapped  systems, the Green's function has no singularities in the 4D space $(\omega,{\bf p})$. That is why the integral \eqref{MasslessTopInvariantSM}  over infinitesimal  surface $\sigma$   vanishes,  $N'(\mu\rightarrow 0)\rightarrow 0$, see Fig. \ref{Crossover}.

However, now we are able to use the other 3D surface $\sigma$   in the integral 
 \eqref{MasslessTopInvariantSM}. The option for $\sigma$ is the momentum slice $p_0=0$, i.e. the integration is over the whole 3D momentum space ${\bf p}$ at fixed $p_0=0$. This option is not invariant  under  Lorentz transformation, which  leads to rotation of the 3D subspace $\sigma$ in the 4D space, but the topological invariants are robust to rotation and do not depend on the choice of the coordinate system.

We shall use the  slice $p_0=0$, i.e.  the Green's function at zero frequency:
 \begin{equation}
S(p_0=0,{\bf p})=  \frac{Z(0,{\bf p}^2)}{-i\tau_2  {\mbox{\boldmath$\sigma$}} \cdot {\bf p}+ M(0,{\bf p}^2)}\,.
\label{eq:calS(p)}
\end{equation}
The propagator at $p_0=0$ has all the properties of a free-fermion Hamiltonian, whose topology was discussed in Ref. \cite{Kitaev2009}, but it emerges in  interacting systems and thus takes into account the interaction (see \cite{Haldane2004,Volovik2009b}).

In  the insulating state of SM, the topological invariant \eqref{MasslessTopInvariantSM1} with $\sigma$ being the slice $p_0=0$  is zero. 
The non-zero topological invariant, which is relevant  for these types of insulators, is the symmetry protected invariant  \cite{Volovik2009b}:
\begin{equation}
\tilde N = {e_{ijk}\over{24\pi^2}} ~
{\bf tr}\left[\tau_2 \int_{p_0=0}   d^3p 
~ {\cal S}\partial_{p_i} S^{-1}
S\partial_{p_j} S^{-1}S\partial_{p_k}  S^{-1}\right].
\label{tildeN}
\end{equation} 
For SM  the matrix  $\tau_2=\gamma_5\gamma^0$  according to \eqref{eq:DiracMatrices}.
This matrix commutes with the Green's function at $p_0=0$, which makes \eqref{tildeN} the topological invariant.

Conventional massive Dirac fermions  with $n_L=n_R=1$ and with momentum independent mass term $M(p^2)=M$  have non-zero topological charge
\begin{equation}
\tilde N= {\rm sign}(M)
\,.
\label{eq:DiracInvariants}
\end{equation}
However, the space of the Green's function of free Dirac fermions is non-compact: $S$ has different asymptotes at $|{\bf p}|\rightarrow \infty $ for different directions of momentum ${\bf p}$.   As a result, the topological charge of free Dirac fermions is ill-defined  and 
even could be fractional. We shall see that it acquires intermediate values between the charges of the vacua with compact   Green's function (see Fig. 2). On the marginal behavior of free Dirac fermions see Refs. \cite{Haldane1988,Schnyder2008,Volovik2003,Volovik2009b}.

\section{Quantum phase transitions}
\label{QuantumPhaseTransitions}

 The interaction may essentially modify the Green's function so that its space becomes compact and thus the topological invariant becomes well defined. As an illustration consider fermions with 
mass function 
 \begin{equation}
M(p^2)=M_0+ M_2p^2 ~.
\label{eq:modified_mass}
\end{equation}
The inverse Green's function approaches infinity for any direction of ${\bf p}$, that is why the space of Green's function is compact.
 For the mass spectrum in \eqref{eq:modified_mass} one has for $M_2>0$: 
\begin{equation}
\tilde N =0~~, ~~M_0>0~~~;~~~   \tilde N =-2~~, ~~M_0<0\,.
\label{3DTopInvariant_modified}
\end{equation} 
Vacuum states with the same symmetry but with different values of the topological invariant $\tilde N$ are shown in Fig. 2 in the plane  $(M_0,M_2)$. The lines of quantum  phase transition (QPT) separate  the trivial insulating vacuum with $\tilde N=0$ and the topological insulators with $\tilde N=\pm 2$.  

\begin{figure}[top]
\centerline{\includegraphics[width=1.0\linewidth]{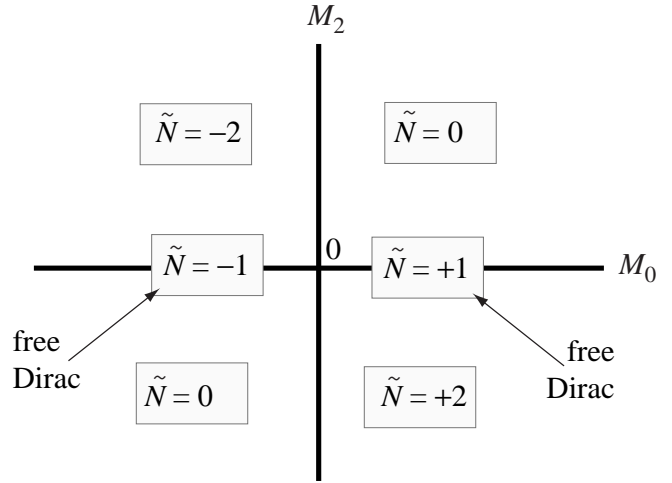}}
\label{QPT} 
  \caption{Fig. 2.  Phase diagram of topological states of the vacuum of  single Dirac fermionic field with mass function $M(p^2)=M_0+ M_2p^2$ in the plane $(M_0,M_2)$. States on the line 
  $M_2=0$ correspond to the vacua of non-interacting Dirac field, whose Green's function space is non-compact. Topological charge of the free Dirac fermions  is intermediate between charges of compact states.
The line $M_2=0$ separates the states with different asymptotic behavior  at infinity:
$S^{-1}(0,{\bf p}) \sim \pm  {\bf p}^2 $. 
 The line $M_0=0$ marks the topological quantum phase transition (QPT) between the topologically trivial insulator with $\tilde N=0$ and topological insulators with $\tilde N=\pm 2$. The intermediate states -- the vacua on the line of this QPT -- are gapless. 
 }
\end{figure}

States on the horizontal axis (line  $M_2=0$) correspond to the vacuum of free Dirac fermions, whose Green's function space is non-compact. Topological charge $\tilde N$ of the free Dirac fermions  is intermediate between charges of compact states with $M_2\neq 0$. The line $M_2=0$ separates the states with different asymptotic behavior  at infinity:
$S^{-1}(0,{\bf p}) \rightarrow \pm  {\bf p}^2$, while fermions remain gapped even at the transition line. 

\begin{figure}[top]
\centerline{\includegraphics[width=1.0\linewidth]{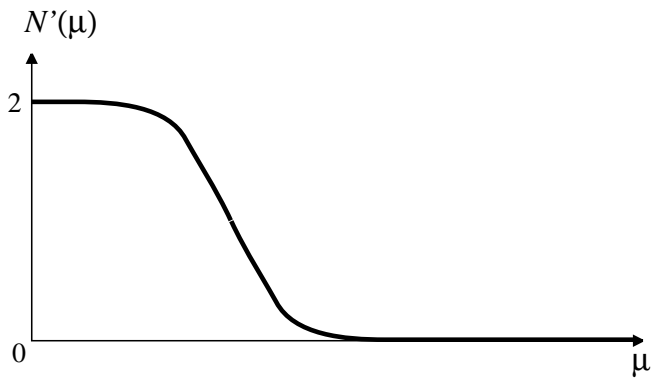}}
  \caption{Fig. 3. Integral \eqref{MasslessTopInvariantSM} as function of running energy $\mu$ in case when symmetry $K$ emerges at small energy. }
 \label{Crossover2} 
\end{figure}

{\it Intermediate gapless states}:   The vertical axis (line $M_0=0$) separates the states with the same asymptote of the Green's function at infinity. The abrupt change of the topological charge across the line, $\Delta \tilde N=2$, with fixed asymptote shows that one cannot cross the transition  line adiabatically. This means that all the intermediate states on the line of this  QPT  are necessarily gapless. We know that the intermediate state between the free Dirac vacua with opposite mass parameter 
 $M$ is massless, but this is applicable not only to the vacuum of free Dirac field but to any state on the line of quantum phase transition. Let us consider the intermediate state at $M_0=0$ and $M_2\neq 0$.  
 There is no symmetry which could protect the gaplessness of this state: the integral $N'$ in \eqref{MasslessTopInvariantSM} which is responsible for massless fermions is not the topological invariant for $M_2\neq 0$ and thus depends on the energy scale $\mu$.
Fig. \ref{Crossover2} demonstrates this dependence. The most striking feature is that $N'(\mu)$ approaches  the integer value when $\mu \rightarrow 0$. This looks as if  the symmetry $K$ emerges at small $p^2$. This happens because we approach the line of topological QPT, at which fermions necessarily become gapless.

In principle, it is not excluded that in a similar way the symmetry $K$ of SM  emerges when $\mu$ decreases below some ultraviolet energy scale $E_{\rm uv}$, and in the region $E_{\rm uv}\gg \mu \gg E_{\rm ew}$ fermions are effectively massless. Then below the electroweak scale, $\mu < E_{\rm ew}$, the symmetry $K$ is violated again leading to massive particles.  Such reentrant violation of symmetry $K$  is possible if the vacuum of our Universe lives on the line of QPT. The reason why nature prefers the critical line of QPT may be that the gapless states on the line are able to accommodate more entropy than the gapped states. 

{\it Fermion zero modes}: 
The gaplessness of the intermediate state leads also to the other related phenomenon. The two-dimensional interface (brane), which separates two domains with different $\tilde N$, contains fermion zero modes,  2+1 massless fermions. The number of zero modes is determined by the difference $\Delta \tilde N$ between the topological charges of the vacua on two sides of the interface   \cite{Volovik2009,Volovik2009b}. This is similar to the index theorem for the number of fermion zero modes on cosmic strings, which is determined by the  topological winding number of the string \cite{JackiwRossi1981}. However, while  topological  charge of string results from spontaneously broken symmetry which determines  topology of defects {\it in real space}, the interface between two vacua is determined by topology {\it in momentum space}.

Since the integrand in \eqref{tildeN} does not obey Lorentz invariance, Eq. \eqref{tildeN}  is applicable also to non-relativistic systems: to superfluid $^3$He-B \cite{SalomaaVolovik1988,Volovik2009b}; for condensed matter Hamiltonians discussed in Refs. \cite{Schnyder2009b} and \cite{Sato2009}; and for many other fully gapped systems.
 In case of lattice models of quantum field theories or for topological insulators in crystals \cite{Volkov1981,VolkovPankratov1985}, the integral is over the 3D  Brillouin zone.
 In particular,  the phase diagram in Fig. 2 is
applicable to superfluid $^3$He-B where the mass term has just the form
$M(0,{\bf p}^2)=M_0+ M_2{\bf p}^2$  \cite{VollhardtWolfle,Volovik2003}. 
Here the parameter  $M_2=1/2m^*$ is related to the effective mass $m^*$ in normal Fermi-liquid; parameter $-M_0$ plays the role of the chemical potential of $^3$He atoms; matrices $\mbox{\boldmath$\tau$}$ are Bogoliubov-Nambu matrices in particle-hole space; and the role of speed of light $c$ is played by the pair-breaking velocity (in both cases we set $c=1$).

\section{Discussion}

Vacuum of SM is a topological medium. Both known states of the quantum vacuum of SM have non-trivial topology.  The insulating state is described by nonzero value of topological invariant $\tilde N$ in \eqref{tildeN}, while  the semi-metal state --  by topological invariant $N'$ in \eqref{MasslessTopInvariantSM}. Both invariants are supported by symmetry.

\begin{figure}[top]
\centerline{\includegraphics[width=1.0\linewidth]{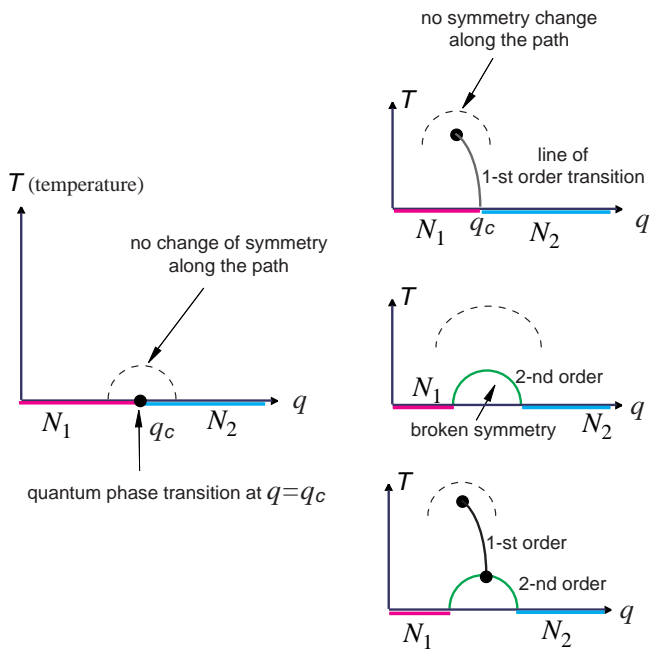}}
  \caption{Fig. 4. Quantum phase transitions (QPT) governed by topology. {\it Left}:  Point of topological quantum phase transition in the plane $(q,T)$, where $q$ is some parameter of the system
  (see e.g. \cite{Volovik2007}). At $T=0$, the vacua are topologically  not equivalent and cannot be adiabatically connected across the transition point $q_c$. They are described either by  different topological invariants, or by different values of the same topological invariant. However, states at $q<q_c$ and $q>q_c$ have the same symmetry and thus can be continuously connected by the path at $T\neq 0$ around the point. Examples are Lifshitz transitions; plateau transitions in quantum Hall effect; semi-metal -- insulator transition  in systems of $^3$He-A type;  transition between the trivial insulator and topological insulator in systems of $^3$He-B type, see Fig. 2;  possibly the confinement-deconfinement transition in QCD and topological quantum phase transitions in SM, where the relevant parameter $q$  may be played by the ratio of high-energy cut-off scales $q=E_{\rm UV}/E_{\rm c}$ in Ref. \cite{KlinkhamerVolovik2005a}. {\it Right}: topological QPT can be interrupted by thermodynamic phase transitions of the 1-st or/and 2-nd order. This type of behavior may occur in QCD.}
 \label{QPTransitions} 
\end{figure}

 Momentum space topology suggests a number of possible quantum phase transitions in the quantum vacuum of SM (see Fig. \ref{QPTransitions}). 
The transition between the semimetal gapless state and the fully gapped insulating state of  the vacuum  is one of them. Condensed matter examples demonstrate that QPT may occur without symmetry breaking, as a purely topological QPT  \cite{Volovik1992b,KlinkhamerVolovik2005b,Volovik2007,GurarieRadzihovsky2007}.
In principle, this is possible in SM too. The symmetry $K$ in the topological invariant $N'$ in \eqref{MasslessTopInvariantSM}, which protects massless fermions in the semimetal state of SM, may be emergent as we discussed in Sec. \ref{QuantumPhaseTransitions}. In this case the electroweak transition or crossover would correspond to QPT in  Fig. 4 {\it right top} or Fig. 4 {\it left}  correspondingly.

There can be also quantum phase transitions  between the insulating states: these states may have the same symmetry, but  different values of topological charge $\tilde N$ in \eqref{tildeN} (see Fig. 2). 
In SM, such states would correspond to different configurations of mass matrices \cite{Jarlskog2005}, which have different values of invariant $\tilde N$. According to \eqref{eq:DiracInvariants} such transition occurs when one of the eigen values of the mass matrix crosses zero. This may occur in the neutrino sector (for example, from the Mikheev-Smirnov-Wolfenstein  type  effects due to the interactions of neutrinos and ambient matter \cite{KlinkhamerComment}). The variety of transitions is enhanced if the fermions of the 4-th generation are added (discussion of the present status of the fourth generation fermions in SM see in \cite{Holdom2009}).

It is not excluded that the phenomenon of quark confinement in quantum chromodynamics is also one of the manifestations of the nontrivial topology of the vacuum medium, which emerges due to interaction of fermions with a non-Abelian gauge field. There are some indications that the phenomenon of  confinement is related to the anomalous infrared properties of the Green's function. As first argued by Gribov \cite{Gribov1978},  a perturbative pole of the gluon propagator is converted into a zero at the vanishing momentum
(see recent papers \cite{Chernodub2008,Burgio2009}).
In principle it is still not excluded by lattice simulations \cite{Bowman2004,Bowman2005}
that the same happens with quark propagator. In condensed matter the conversion of pole to zero has been also discussed, see e.g.  \cite{Volovik2007,Chernodub2008,FaridTsvelik2009} and references therein. This conversion occurs as a quantum phase transition, and one may expect that the confinement and deconfinement states of quantum vacuum are also separated
by a similar quantum phase transition related to topology of the Green's function. In this case 
if the  lines of thermodynamic phase transitions are wiped out from the QCD phase diagram in Ref.\cite{Ruester2005} (whose simplified version is in Fig. 4 {\it right}), then what is left would be the topological QPT between the QCD vacua with and without confinement. 
 
Topological analysis of quantum vacua in terms of Green's function becomes even more important if SM is not a fundamental theory, but is an effective theory, where all the gauge symmetries emerge only at low energy.  In this case the Green's function is the general $32n_g\times 32 n_g$ matrix, which does not split into blocks. Its elements may be connected  by some  discrete symmetries of the underlying physics, such as $Z_2$ and $Z_4$ symmetries $K$ discussed in Sec. \ref{TopSemi-metal}.  These discrete symmetries give rise to topological invariant  $N'$ in \eqref{MasslessTopInvariantSM} which generates emergent chiral fermions at low energy and also serves as a source of emergent gauge groups  \cite{Volovik2003}.

If the Lorentz symmetry and CPT are also emergent 
(on the present status of bounds on violation of these symmetries see  \cite{Gonzalez-Mestres2009} and references therein), the other types of quantum phase transition are possible  in SM. Among them the splitting of Fermi points and formation of Fermi surfaces with non-zero global  topological charge $N$ in \eqref{MasslessTopInvariantSM1}  \cite{KlinkhamerVolovik2005b,Volovik2007}. Such transitions lead in particular to induced Chern-Simons terms in effective action with parameters determined by splitting. In 3D condensed matter systems these parameters correspond to a non-quantized part of the intrinsic Hall conductivity  \cite{KlinkhamerVolovik2005b,Haldane2004}. Reentrant violation of Lorentz symmetry which occurs at low energy   leads to formation of exotic massless fermions with nontrival momentum space topology: fermions with quadratic dispersion at low energy and  semi-Dirac fermions with linear dispersion in one direction and quadratic dispersion in the other  \cite{Volovik2001,Volovik2003}. Examples of such fermions  in  condensed matter  are in Refs.  
\cite{Volovik2001,Volovik2007,Dietl-Piechon-Montambaux2008,Banerjee2009}.

I thank Maxim Chernodub and Frans Klinkhamer for numerous fruitful discussions. This work is supported in part  by the Academy of Finland, Centers of Excellence Program 2006--2011,
the Russian Foundation for Basic Research (grant 06--02--16002--a),
and the Khalatnikov--Starobinsky leading scientific school (grant 4899.2008.2).


\end{document}